\documentclass{article}[10pt]

\usepackage[utf8]{inputenc} %unicode support

\usepackage{graphicx}
\usepackage{multirow}

\title{Linking de novo assembly results with long DNA reads by dnaasm-link application}

\author{Wiktor Kuśmirek$^{1}$\footnote{W.Kusmirek@ii.pw.edu.pl} \and Wiktor Franus$^{1}$ \and Robert Nowak$^{1}$\\
  Warsaw University of Technology, Institute of Computer Science}
\date{}

\begin{document}

\maketitle

\begin{abstract} % abstract
Background

Currently, third-generation sequencing techniques, which allow to obtain much longer DNA reads compared to the next-generation sequencing technologies, are becoming more and more popular.
There are many possibilities to combine data from next-generation and third-generation sequencing.

Results

Herein, we present a new application called dnaasm-link for linking contigs, a result of \textit{de novo} assembly of second-generation sequencing data, with long DNA reads.
Our tool includes an integrated module to fill gaps with a suitable fragment of appropriate long DNA read, which improves the consistency of the resulting DNA sequences.
This feature is very important, in particular for complex DNA regions, as presented in the paper.
Finally, our implementation outperforms other state-of-the-art tools in terms of speed and memory requirements, which may enable the usage of the presented application for organisms with a large genome,
which is not possible in~existing applications.

Conclusions

The presented application has many advantages as (i) significant memory optimization and reduction of computation time (ii) filling the gaps through the appropriate fragment of a specified long DNA read (iii) reducing number of spanned and unspanned gaps in the existing genome drafts.

The application is freely available to all users under GNU Library or Lesser General Public License version 3.0 (LGPLv3).
The demo application, docker image and source code are available at http://dnaasm.sourceforge.net.

\end{abstract}

%%%%%%%%%%%%%%%%%%%%%%%%%%%%%%%%%%%%%%%%%%%%%%
%%                                          %%
%% The keywords begin here                  %%
%%                                          %%
%% Put each keyword in separate \kwd{}.     %%
%%                                          %%
%%%%%%%%%%%%%%%%%%%%%%%%%%%%%%%%%%%%%%%%%%%%%%

Keywords: {hybrid de~novo assembling}, {next generation sequencing}, {third generation sequencing}, {draft genome sequence}

\section*{Background}

High-throughput sequencing devices, called next-generation sequencers, have provided lots of DNA sequences of various organisms.
However, a very large number of draft genome is still incomplete, for example, in GenBank, 90\% of bacterial genomes are incomplete~\cite{bacterialGenomeSequencing}.
In order to improve the consistency and completeness of draft of reference genomes, which are produced based on short reads obtained by second-generation sequencers,
third-generation long reads sequencing can be used.
Due to this fact, third-generation sequencing technologies are becoming more popular,
for example, in 2018 the human genome assembled \textit{de novo} from only long DNA reads was published~\cite{human-genome-ONT-sequencing}.

Third-generation sequencing allowed to obtain much longer DNA reads compared to the second-generation sequencing technologies.
However, the error rate in long reads from third-generation devices compared to short DNA reads from second-generation sequencers is significantly higher~\cite{pacbio-sequencing, nanopore-sequencing}.
Moreover, the cost per sample of third-generation sequencing is higher than the second-generation sequencing~\cite{cost-per-sample-sequencing}.

An obvious concept of using both types of reads in \textit{de novo} assembly called hybrid assembly is currently explored~\cite{apple-hybrid-assembly, herbal-plant-hybrid-assembly}.
There are many possibilities to combine data from second-generation sequencing and third-generation sequencing.
The four most popular are listed below.
\begin{enumerate}
\item Long DNA reads could be mapped directly to the de Bruijn graph, which is built from short DNA reads.
  Then, dedicated algorithms allow us to resolve some ambiguity in the de Bruijn graph, which can improve the consistency of the resulting DNA sequences.
  Such an approach is implemented in some \textit{de novo} DNA assemblers for second-generation reads, e.g. Velvet~\cite{velvet}, ABySS~\cite{abyss}, SPAdes~\cite{spades}.
\item Long DNA reads could be \textit{de novo} assembled with dedicated assemblers, e.g. Canu~\cite{canu}, Falcon~\cite{falcon}, miniasm~\cite{miniasm}.
  Then, created DNA sequences can be improved in terms of quality by mapping short DNA reads and correcting assembling errors, by Pilon~\cite{pilon} or quiver~\cite{quiver} applications.
\item Short DNA reads could be used to correct long DNA reads, for example, with CoLoRMap~\cite{colormap} or Nanocorr~\cite{nanocorr} tools.
  Then, long and corrected DNA reads could be assembled with assemblers for third-generation sequencing data (as depicted in the previous point).
\item Short DNA reads could be \textit{de novo} assembled using assemblers dedicated for second-generation sequencing data (as depicted in point 1).
  Then, long DNA reads could be used to link the resultant DNA sequences (contigs), for example, with LINKS~\cite{links} or SSPACE-LongRead~\cite{sspace-longread} applications.
\end{enumerate}

In this paper, we present a new application called \textit{dnaasm-link} for combining output of \textit{de novo} assembler with long DNA reads (point 4 of the previous list).
Our software contains a module for filling the gaps between contigs with a specified sequence from an appropriate long DNA read.
This feature is very important, in particular for complex DNA regions.
What is more, our method has several times shorter calculation time as well as several times lower memory requirements in comparison to other tools.
Significant memory optimization and reduction of computation time may enable the usage of the presented application for organisms with a large genome,
which is not possible in~existing applications.

The presented algorithm was implemented as a new extension of the dnaasm assembler~\cite{bmc2018dnaasm},
the demo application, docker image and source code are available at project homepage http://dnaasm.sourceforge.net.

\section*{Methods}

The presented algorithm efficiently finds and joins the adjacent contigs using long reads.
The contigs are produced by \textit{de novo} DNA assembler from short and high quality reads from second-generation sequencers.
In our approach the contigs are created by de Bruijn graph algorithm implemented in dnaasm assembler~\cite{bmc2018dnaasm}.
The new algorithm, called dnaasm-link, checks which contigs have a~sub-sequence similar to a~sub-sequence in long reads,
then adjacent contigs are found,
the distance between contigs is calculated and the gap is filled with a~sequence from the appropriate long DNA read.
The presented approach and the implementation details are described below.

\subsection*{Finding adjacent contigs}

The algorithm to find adjacent contigs uses k-mers similarity.
This algorithm consists of several stages.

\begin{figure}[h!]
  \includegraphics[scale=0.4,angle=270]{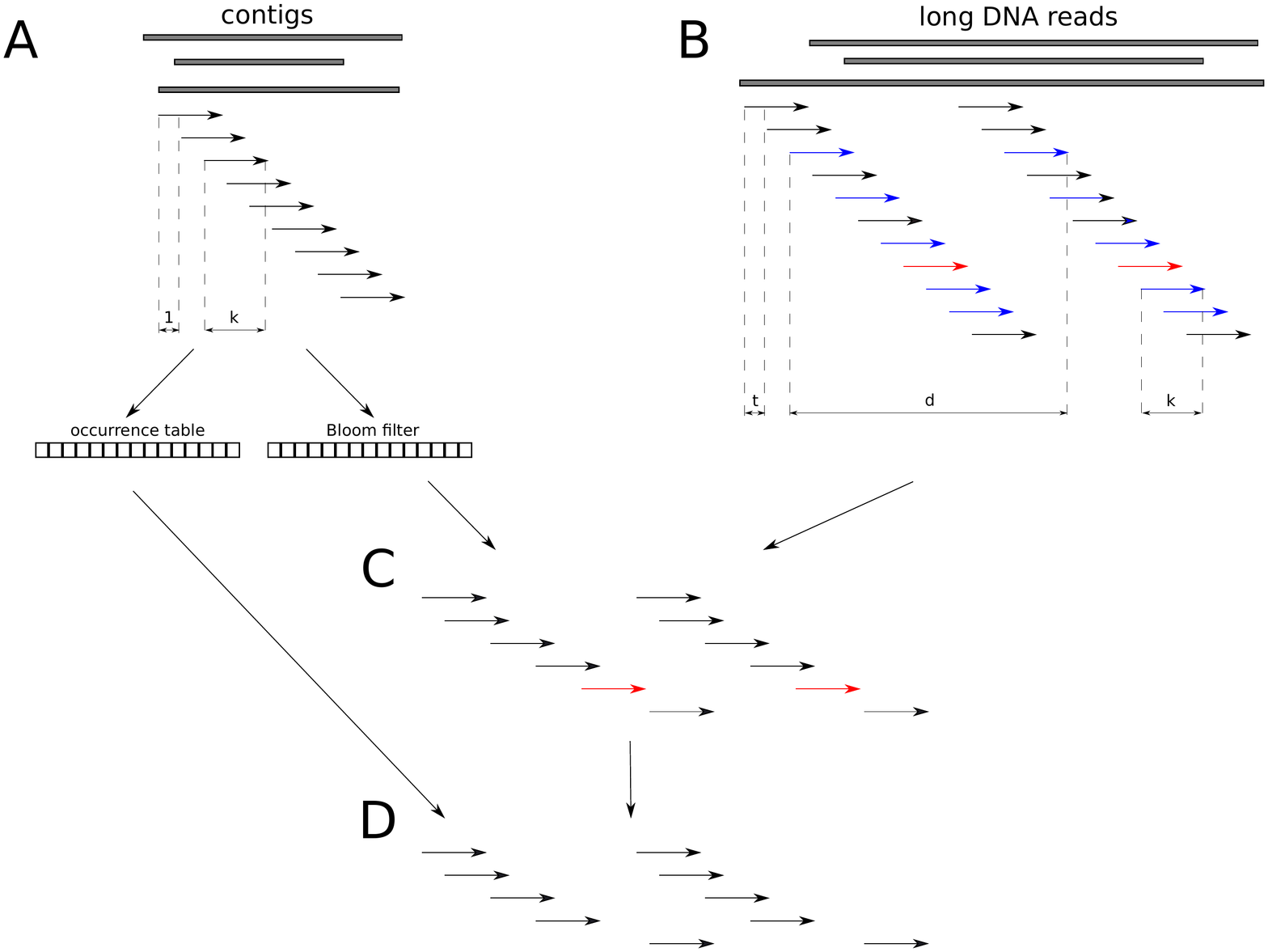}
  \caption{An exemplary process of generating and filtering k-mers pairs from long DNA reads.
    (A) Firstly, the Bloom filter and an array, containing the number of occurrences of each k-mer, are build based on the k-spectrum generated from the input set of contigs.
    (B) From each long DNA read, a set of k-mers pairs (k-mer length equal to $k$) is generated, with a distance between the beginning of the first k-mer and the end of the second one equal to $d$ and a sliding step equal to $t$.
    (C) The input set of k-mers pairs is filtered with the Bloom filter - some pairs are discarded (blue arrows).
    (D) Resultant set of k-mers pairs after the second filtering process.
    It is worth noting that the resulting set of k-mers pairs (D) is very limited in relation to the generated set of k-mers pairs (B) due to errors in long DNA reads and repetitive regions of the investigated genome.}
  \label{fig:algorithm-before-graph-building}
\end{figure}

Firstly, a set of k-mers is generated from the input set of contigs, each of them being inserted into the Bloom filter~\cite{bloom-filter}.
The length of analyzed k-mers (the value of parameter $k$) can be set by the user based on the error rate of long DNA reads - the higher error rate, the lower $k$ value.
The default value is 15.
This step is depicted in~Fig.~\ref{fig:algorithm-before-graph-building}A.

Secondly, a~set of long DNA reads is proceeded - the set of k-mers pairs with the distance $d$ is generated.
The default value $d$ is $4000$.
It should be mentioned, that we do not generate a~full k-spectrum here, we rather use the step value $t$, set by default to $2$.
This step is depicted in~Fig.~\ref{fig:algorithm-before-graph-building}B.
The pairs in which both k-mers are in the generated previously Bloom filter, are processed further, as depicted in~Fig.~\ref{fig:algorithm-before-graph-building}C.

Thirdly, a set of unique k-mers is determined.
This process consists in counting the number of instances of a given k-mer in the input set of contigs.
K-mers which occur more than once are treated as non-unique.
All pairs of k-mers containing at least one non-unique k-mer are removed from further considerations,
as depicted in~Fig.~\ref{fig:algorithm-before-graph-building}D.

Next, the connection graph is built.
This graph is composed of vertices that represent contigs, and edges that represent connections between contigs derived from pairs of k-mers from long DNA reads.
Each edge contains three parameters that define the strength of the specified connection.
These parameters are:
\begin{itemize}
  \item the number of connections between a given pair of contigs defined as the number of k-mers pairs;
  \item the number of connections between a given pair of contigs defined as the number of DNA reads;
  \item the number of connections between a given pair of contigs defined as the number of DNA reads, where specified DNA read is taken into consideration if number of k-mers pairs in this read is greater than the threshold value specified by the user.
\end{itemize}
After building the connection graph a set of filters is applied to remove some edges representing connections.
Filters remove the edges where at least one of the three numbers mentioned above is lower than the corresponding thresholds set by the user.

Finally, the process of generating the resulting set of scaffolds from the connection graph is performed.
The process respectively considers not processed previously vertices of the graph, starting from the vertex that represents the longest contig.
Then, the considered contig is expanded to the left and right.
During the expansion, two types of situations can occur:
(i) the specified contig is connected only with a single vertex in contig graph, then, considered contigs are joined;
(ii) the specified contig is connected with more than a single vertex. In this situation, a vertex with the largest number of connecting pairs of k-mers is preferred.
All vertices used in expansion are marked as used and are not taken into consideration in the next iteration of the algorithm.
The process is repeated till all vertices are processed.

\subsection*{Gap filling algorithm}

During scaffolds generation two contigs may overlap, then a single 'N' sign is inserted between them.
However, the contigs may be separated by a gap and the final step of the presented algorithm aims
to estimate the gap size and to fill it with an appropriate fragment of long DNA read.
Since both the distance between paired k-mers and the coordinates of those k-mers on
contigs are known, the estimated length of each gap is calculated.
In the same manner, having known the offset of each k-mer pair extracted from the long read,
it is possible to determine the offset of a sub-sequence of a read corresponding to each of the gaps in scaffolds.
Contigs are covered by multiple error-containing reads, and consequently, multiple different gap sequences may be generated.
In the presented application,
a gap sequence is taken directly from the read which covers the considered contigs with the greatest number of k-mer pairs.

\subsection*{Implementation}

\begin{figure}[h!]
  \includegraphics[scale=0.4,angle=270]{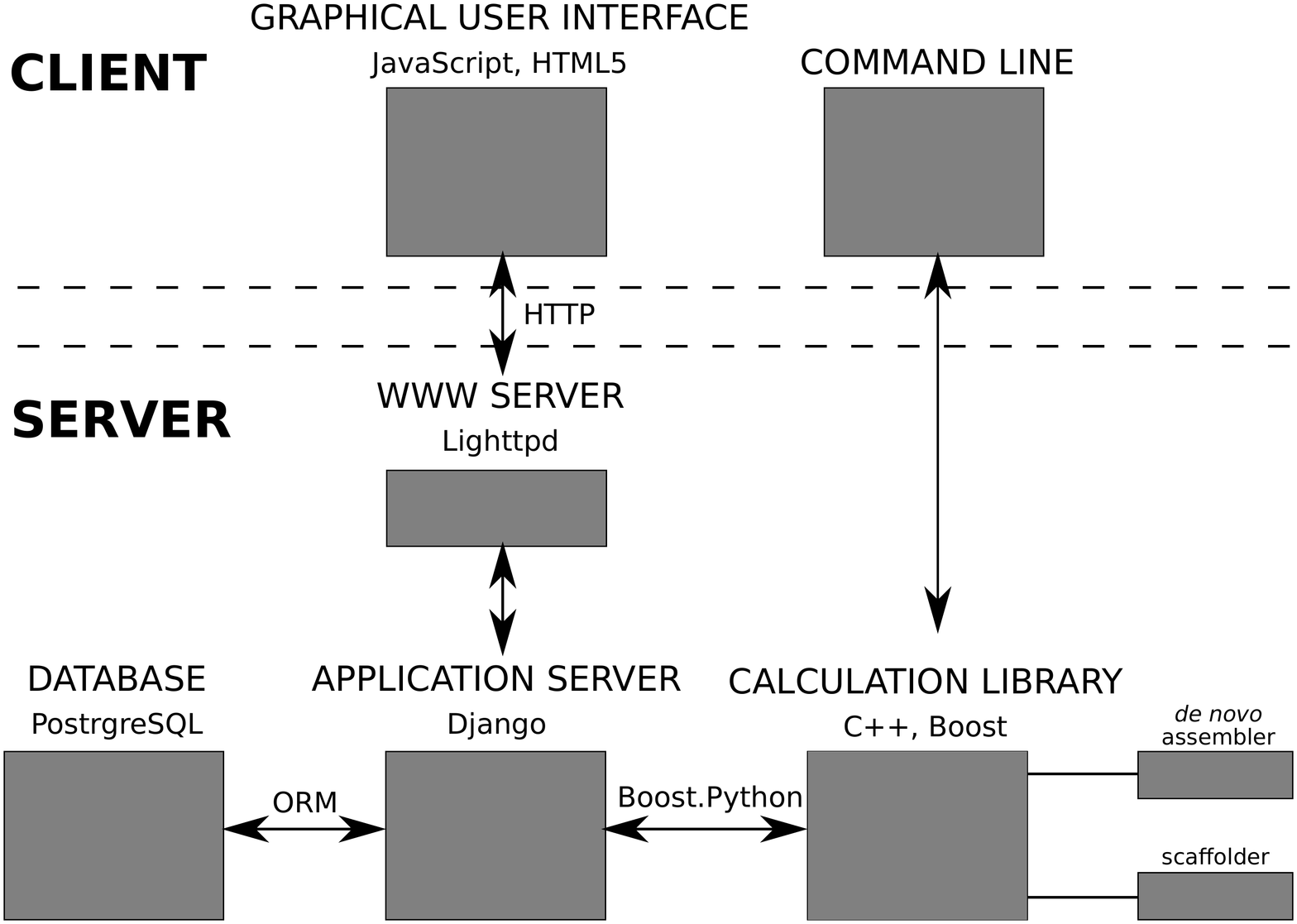}
  \caption{The architecture of dnaasm application. The user can use the application in two ways, through graphical user interface or a~command line.
    Both ways lead to launching the calculation module, in which the presented algorithm is implemented as the shared library. What is more, the calculation module contains an additional shared library in which the \textit{de novo} assembler was implemented beforehand.
    Both the mentioned assembler and the presented dnaasm-link scaffolder can be launched in a very similar and convenient way.}
  \label{fig:architecture}
\end{figure}

The dnaasm application was implemented in the client-server architecture, based on the bioweb framework~\cite{bioweb}.
The dnaasm-link is a new module, deployed as shared library.
In our implementation, we used three programming languages: JavaScript, Python and C++.
Firstly, JavaScript along with HTML5 and AngularJS framework were used to implement graphical user interface (GUI).
Then, Python and Django library were used to implement server side.
Finally, C++ was used to implement the most complex data processing step - the algorithm presented in the work.
Moreover, we used several libraries, like Boost and Google Sparse Hash, to make implementation of our algorithm fast and memory scalable.
The main modules of our software are presented in Fig~\ref{fig:architecture}.

\section*{Results}

Numerical experiments were performed to compare the presented application with the available tools and to indicate the advantages of gap filling in scaffolds using long DNA reads.
Briefly, the first experiment compares the quality of results obtained in the presented method with other tools for hybrid assembly.
The second experiment was carried out on artificially generated data and it indicates the benefits of using both short and long DNA reads over using only the output from second-generation sequencers.
Finally, the calculation time and memory usage were measured.

To evaluate the quality of resultant DNA sequences in experiments we used QUAST~\cite{quast} ver.~4.1.
We compared DNA sequences in terms of:
\begin{itemize}
\item{the number of resultant DNA sequences longer than 1000 bp;}
\item{the number of misassemblies - sum of relocations, translocations, and inversions;}
\item{N50 statistic - the length of the DNA sequence for which the sum of lengths of all sequences of that length or longer is greater than half of an assembly;}
\item{NA50 statistic - the same as N50, but not for all resultant DNA sequences - only for a set of aligned blocks which are results of breaking input DNA sequences at misassembly events;}
\item{the largest DNA sequence;}
\item{the largest alignment - the length of the largest continuous alignment in the resultant DNA sequences;}
\item{the average number of mismatches per 100 kbp;}
\item{the average number of indels per 100 kbp;}
\item{the average number of uncalled bases (N's) per 100 kbp.}
\end{itemize}
Moreover, we used BUSCO~\cite{busco} ver.~2.0 tool to compare the DNA sequence in terms of the number of reconstructed core genes.
As part of this evaluation of the DNA sequences, we have distinguished four groups: (i) complete and single-copy, (ii) complete and duplicated, (iii) fragmented and (iv) missing core genes.
A detailed description of the experiments and the results obtained could be found in the next parts of this section.

\subsection*{Comparison with another tools}

We compared the results of our application with other tools for hybrid assembly that connect the contigs using long reads.
The main objective was comparison in terms of linking the contigs with long DNA reads and filling in the resulting gaps.
For the above experiment we used publicly available data for \textit{Escherichia coli} and \textit{Saccharomyces cerevisiae}.
Both of the datasets on which we worked came from Nanocorr's~\cite{nanocorr} research\footnote{http://schatzlab.cshl.edu/data/nanocorr}, the names of the files are provided in Supplementary materials.
The above files are the result of \textit{de novo} assembly of short DNA reads and the correction of ONT reads by short DNA reads.
Basic parameters of the input set of long DNA reads and contigs are presented in Tab.~\ref{tab:coli_yeast_long_reads}.

\begin{table}[h!]
  \caption{The input set of long DNA reads and contigs characteristic for \textit{E. coli} and \textit{S. cerevisiae} organisms from Nanocorr's research.}
\begin{tiny}
  \begin{tabular}{l|r|r|r|r|r|r|r|r|} 
    \multicolumn{2}{c|}{~} & \textbf{No. of}    & \multirow{2}{*}{\textbf{Sum [Mbp]}} & \multirow{2}{*}{\textbf{N50 [bp]}} & \multirow{2}{*}{\textbf{Max [bp]}} & \textbf{Avg.} & \textbf{Avg.} & \textbf{Avg.} \\
    \multicolumn{2}{c|}{~}  & \textbf{sequences} & ~ & ~ & ~ & \textbf{mis.} & \textbf{indels} & \textbf{N's} \\ \hline
    \multirow{2}{*}{contigs} & \textit{E. coli} & 65 & 4.681 & 176396 & 398301 & 2.32 & 0.17 & 0.00   \\
    ~ & \textit{S. cerevisiae} & 430 & 14.911 & 53444 & 257346 & 85.77 & 8.80 & 0.00  \\ \hline
    \multirow{2}{*}{long reads} & \textit{E. coli} & 59009 & 240.098 & 7471 & 43798 & 180.75 & 181.20 & 0.00   \\
    ~ & \textit{S. cerevisiae} & 88218 & 526.589 & 9189 & 72879 & 360.98 & 171.80 & 5.06  \\ \hline
  \end{tabular}
\end{tiny}
  \label{tab:coli_yeast_long_reads}
\end{table}

We compare our approach with two state-of-the-art tools used to join contigs into scaffolds with long reads: LINKS~\cite{links} ver. 1.8.5 and SSPACE-LongRead~\cite{sspace-longread} ver. 1.1.0.
What is more, in our research we also used some scaffolders for short, paired DNA reads - paired-end tags (PETs) or mate-pairs (MPs).
In order to run scaffolders for short DNA reads on long DNA reads dataset, we used the Fast-SG~\cite{fast-sg} tool.
This application allows us to generate a~set of paired DNA reads from subsequent long DNA reads and to map them to the preassembled contigs.
Then, this set of mapped, short DNA reads along with contigs were used as an input for scaffolders dedicated to short reads: OPERA-LG~\cite{opera-lg} ver. 2.0.6, BOSS~\cite{boss} and ScaffMatch~\cite{scaffmatch} ver. 0.9.0.
Parameter values for applications and the appropriate commands are provided in Supplementary materials,
while the results of the evaluation are presented in Tab.~\ref{tab:schatzlab-coli-yeast-linkage-statistics} and Tab.~\ref{tab:schatzlab-coli-yeast-core-genes}.

\begin{table}[h!]
  \caption{Evaluation of dnaasm-link application in comparison to other tools for datasets depicted in Tab.~\ref{tab:coli_yeast_long_reads}.
    The following reference sequences were used to evaluate the results: NC\_000913 for \textit{E. coli} and NC\_001133 ... NC\_001148, NC\_001224 for \textit{S. cerevisiae}.}
  \begin{tiny}
  \begin{tabular}{l|r|r|r|r|r|r|r|r|r|r|}
    \multicolumn{2}{c|}{~} & \textbf{No. of} & \textbf{No. of} & \textbf{N50} & \textbf{NA50} & \textbf{Max} & \textbf{Largest} & \textbf{Avg.} & \textbf{Avg.} & \textbf{Avg.} \\
    \multicolumn{2}{c|}{~} & \textbf{contigs} & \textbf{mis.} & \textbf{[bp]} & \textbf{[bp]} & \textbf{[bp]} & \textbf{algn. [bp]} & \textbf{mis.} & \textbf{indels} & \textbf{N's} \\ \hline
    \multirow{7}{*}{\textit{E. coli}} & NGS contigs & 65 & 9 & 176396 & 164044 & 398301 & 360084 & 2.32 & 0.17 & 0.00 \\
    ~ & SSPACE-LongRead & 32 & 29 & 398301 & 211043 & 1274776 & 564486 & 2.47 & 0.37 & 570.90 \\
    ~ & LINKS & 23 & 19 & 637611 & 235726 & 1146701 & 636452 & 2.36 & 0.39 & 233.43 \\
    ~ & \textbf{dnaasm-link} & \textbf{22} & \textbf{20} & \textbf{746714} & \textbf{219242} & \textbf{1128693} & \textbf{636452} & \textbf{2.36} & \textbf{0.37} & \textbf{212.75} \\
    ~ & Fast-SG + OPERA-LG & 26 & 16 & 349966 & 342146 & 659623 & 658295 & 2.36 & 0.30 & 326.53 \\
    ~ & Fast-SG + BOSS & 60 & 14 & 177523 & 164044 & 611106 & 360084 & 2.32 & 0.17 & 64.79 \\
    ~ & Fast-SG + ScaffMatch & 55 & 18 & 185955 & 177523 & 603113 & 359089 & 2.41 & 0.17 & 139.44 \\ \hline
    \multirow{7}{*}{\textit{S. cerevisiae}} & NGS contigs & 430 & 53 & 53444 & 49075 & 257346 & 249232 & 85.77 & 8.80 & 0.00 \\
    ~ & SSPACE-LongRead & 557 & 105 & 167867 & 126607 & 736874 & 452023 & 95.42 & 11.27 & 3690.74 \\
    ~ & LINKS & 202 & 89 & 202618 & 126598 & 623140 & 416048 & 87.04 & 10.00 & 850.77 \\
    ~ & \textbf{dnaasm-link} & \textbf{190} & \textbf{92} & \textbf{224004} & \textbf{126353} & \textbf{764024} & \textbf{431875} & \textbf{87.28} & \textbf{10.08} & \textbf{861.19} \\
    ~ & Fast-SG + OPERA-LG & 202 & 59 & 180866 & 155226 & 736942 & 451889 & 85.51 & 9.72 & 462.50 \\
    ~ & Fast-SG + BOSS & 369 & 113 & 57097 & 47994 & 257346 & 249232 & 85.77 & 8.80 & 374.16 \\
    ~ & Fast-SG + ScaffMatch & 328 & 144 & 80833 & 51157 & 434320 & 249232 & 85.41 & 8.82 & 489.70 \\ \hline
  \end{tabular}
  \end{tiny}
  \label{tab:schatzlab-coli-yeast-linkage-statistics}
\end{table}

\begin{table}[h!]
  \caption{Comparison of the number of core genes reproduced from datasets depicted in Tab.~\ref{tab:coli_yeast_long_reads}.
    The sets of reference core genes used for evaluation were enterobacteriales\_odb9 and saccharomycetales\_odb9 for \textit{E. coli} and \textit{S. cerevisiae}, respectively.}
  \begin{tiny}
  \begin{tabular}{l|r|r|r|r|r|r|r|r|r|}
    \multicolumn{2}{c|}{~} & \textbf{Complete and} & \textbf{Complete and} & \multirow{2}{*}{\textbf{Fragmented}} & \multirow{2}{*}{\textbf{Missing}} \\
    \multicolumn{2}{c|}{~} & \textbf{single-copy} & \textbf{duplicated} & ~ & ~ \\ \hline
    \multirow{7}{*}{\textit{E. coli}} & NGS contigs & 780 & 0 & 1 & 0 \\
    ~ & SSPACE-LongRead & 619 & 162 & 0 & 0 \\
    ~ & LINKS & 780 & 0 & 1 & 0 \\
    ~ & \textbf{dnaasm-link} & \textbf{780} & \textbf{0} & \textbf{1} & \textbf{0} \\
    ~ & Fast-SG + OPERA-LG & 780 & 0 & 1 & 0 \\
    ~ & Fast-SG + BOSS & 780 & 0 & 1 & 0 \\
    ~ & Fast-SG + ScaffMatch & 780 & 0 & 1 & 0 \\ \hline
    \multirow{7}{*}{\textit{S. cerevisiae}} & NGS contigs & 1657 & 9 & 18 & 27 \\
    ~ & SSPACE-LongRead & 1647 & 27 & 15 & 22 \\
    ~ & LINKS & 1661 & 9 & 14 & 27 \\
    ~ & \textbf{dnaasm-link} & \textbf{1659} & \textbf{10} & \textbf{14} & \textbf{28} \\
    ~ & Fast-SG + OPERA-LG & 1661 & 9 & 16 & 25 \\
    ~ & Fast-SG + BOSS & 1660 & 9 & 18 & 24 \\
    ~ & Fast-SG + ScaffMatch & 1658 & 9 & 12 & 32 \\ \hline
  \end{tabular}
  \end{tiny}
  \label{tab:schatzlab-coli-yeast-core-genes}
\end{table}

Our experiment indicates that the dnaasm-link application gives slightly better results than existing tools in terms of the quantity and quality of the resulting DNA sequences.
What is more, \textit{de novo} assembling process by tools that treat short and long reads differently (LINKS, SSPACE-LongRead, dnaasm-link) gives better results
than converting long reads into short reads to increase sequencing coverage followed by \textit{de novo} assembling.

\subsection*{The impact of adding long DNA reads}

We examined how the combination of short and long DNA reads affects the length and quantity of the resulting DNA sequences.
In this study we used \textit{Saccharomyces cerevisiae} (GenBank NC\_001133 ... NC\_001148, NC\_001224) reference genome.
From this genome we generated nine sets of short DNA reads by the pIRS~\cite{pirs} ver. 1.1.1 application and five sets of long reads by NanoSim~\cite{nanosim} ver. 1.0.0 tool,
where each set had a different depth of coverage.
The details of application used and dataset parameters are provided in Supplementary materials.

\begin{figure}[h!]
  \includegraphics[scale=0.33,angle=0]{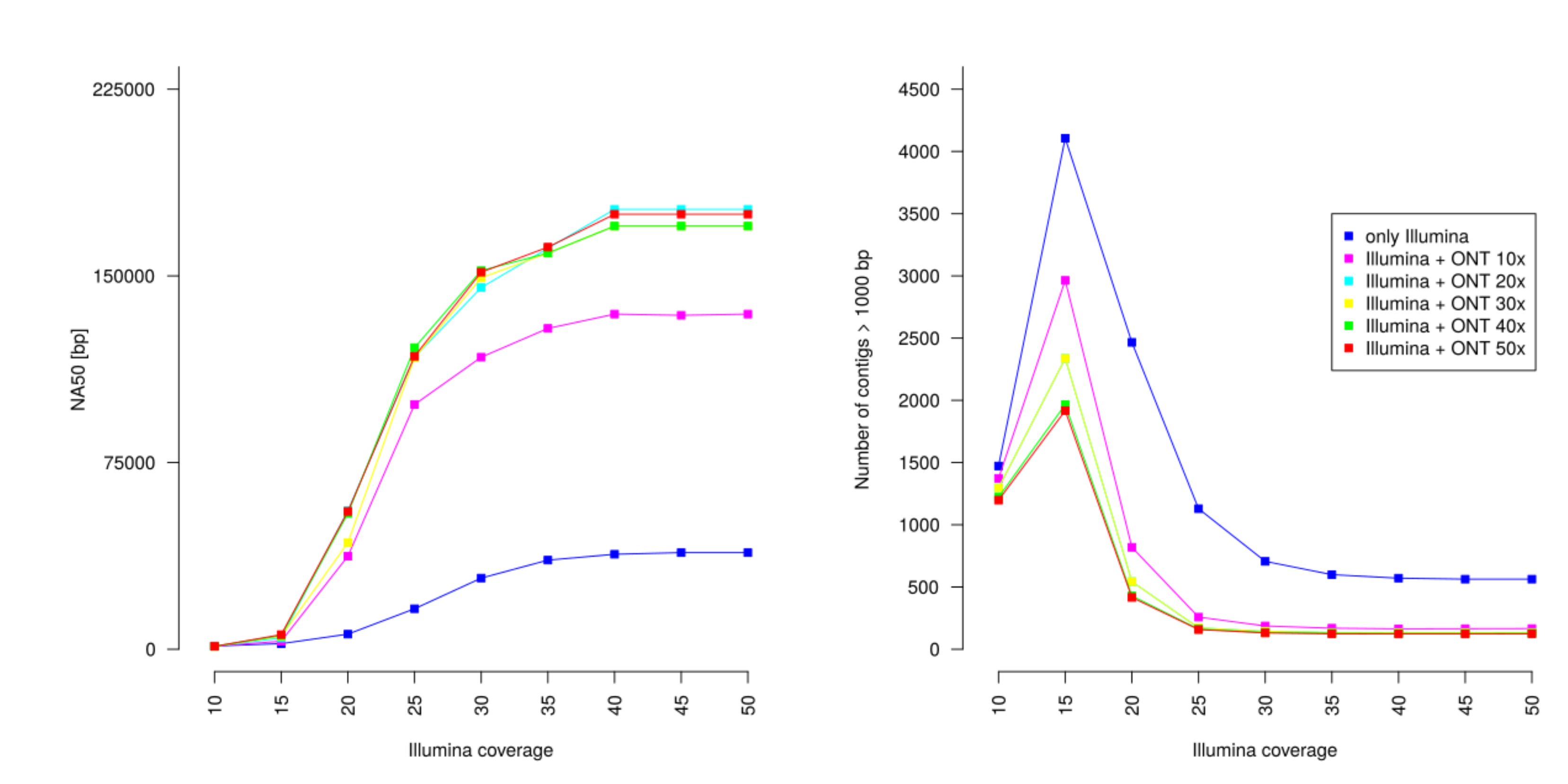}
  \caption{The impact of adding long DNA reads on the number of resultant scaffolds longer than 1000 bp and NA50 statistic. The experiment was carried out on \textit{Saccharomyces cerevisiae} (GenBank NC\_001133 ... NC\_001148, NC\_001224) genome. Firstly, nine sets of short DNA reads and five sets of long DNA reads with different depth of coverage were generated. Then, short reads were \textit{de novo} assembled, and finally, resultant unitigs were linked by long DNA reads. The peak in number for contigs for Illumina coverage equal to 15x is due to the fact that 10x is too small to cover the whole genome.
    After increasing the coverage, the number of contigs increases at the beginning,
    because the whole genome is covered, but with small gaps.
    It is worth mentioning that a~greater depth of coverage does not increase a~number of covered gaps in the results,
    as all the gaps are caused by the complex DNA region and not the lack of coverage.}
  \label{fig:mixing_illumina_ont}
\end{figure}

The generated short reads were \textit{de novo} assembled by ABySS ver. 2.0.1, then contigs were linked using long reads.
The results, presented in Fig~\ref{fig:mixing_illumina_ont}, prove that
combining long DNA reads with short ones can significantly increase the consistency of the resultant assemblies by reducing the final number of scaffolds.
Moreover, increasing the coverage of any sequencing technology above the certain level does not improve the results any more.

Next, we investigate how the use of long DNA reads affects the reconstruction of complex DNA structures such as long tandem repeats.
We compare our method to a~technique where gaps are filled with short DNA reads.
In this experiment we generated an~input set of reads for two organisms: \textit{Escherichia coli} (GenBank NC\_000913) and \textit{Saccharomyces cerevisiae} (GenBank NC\_001133 ... NC\_001148, NC\_001224).
We used the same applications, ie. pIRS and NanoSim as before, whose parameters are provided in Supplementary materials.
The short reads were \textit{de novo} assembled by ABySS~\cite{abyss}.
Next, we link contigs with long DNA reads by dnaasm-link tool in two modes: with and without gap filling.
Then, the scaffolds produced by dnaasm-link without gap filling were treated by three tools for filling gaps with short DNA reads:
GapFiller~\cite{gapfiller} ver. 1.10.0 , Sealer~\cite{sealer} ver. 1.9.0 and SOAPdenovo2~GapCloser~\cite{soapdenovo2-gapcloser} ver. 1.12.0.
Finally, we compared a number of detected tandem repeats by Tandem repeats finder application~\cite{trf}.
The mentioned application was also launched on the reference genomes, to determine ground truth data for this study.
The results presented in Tab.~\ref{tab:gap-filling-efficiency-tandem-repeats} depict the advantage of gap filling by dnaasm-link over other existing methods.

\begin{table}[h!]
  \caption{Tandem repeat reconstruction efficiency. The table presents all tandem repeats in the \textit{E. coli} and \textit{S. cerevisiae} reference genomes.
    In the presented table '+' signs mean the correct reproduction of the specified repetitive fragment,
    '-' signs the lack of correct reconstruction.
    The presented results indicate that the usage of long DNA reads by dnaasm-link tool allows to  reconstruct some of tandem repeats.}
  \begin{tiny}
  \begin{tabular}{l|r|r|r|r|r|r|r|r|r|r|r|}
    \multirow{3}{*}{\textbf{}} & \textbf{Motif} & \textbf{Num of} & \textbf{NGS} & \textbf{dnaasm-link} & \multicolumn{3}{|c|}{\textbf{dnaasm-link without gap filling}} & \textbf{dnaasm-link} \\
    ~ & \textbf{len. [bp]} & \textbf{repet.} & \textbf{unitigs} & \textbf{without gap fill.} & \textbf{+ GapFiller} & \textbf{+ Sealer} & \textbf{+ GapCloser} & \textbf{with gap fill.} \\ \hline
    \multirow{6}{*}{\textit{E. coli}} & 181 & 3.0 & - & - & - & - & - & - \\
    ~ & 181 & 2.3 & - & - & - & - & - & - \\
    ~ & 178 & 1.9 & - & - & + & - & - & + \\
    ~ & 226 & 2.0 & - & - & - & - & - & + \\
    ~ & 113 & 2.7 & - & - & - & - & - & + \\
    ~ & 226 & 1.9 & - & - & - & - & - & - \\
    ~ & 200 & 2.0 & - & - & - & - & - & + \\ \hline
    \multirow{14}{*}{\textit{S. cerevisiae}} & 135 & 1.9 & - & - & - & - & - & - \\
    ~ & 135 & 1.9 & - & - & - & - & - & - \\
    ~ & 135 & 3.1 & - & - & - & - & - & - \\
    ~ & 135 & 3.1 & - & - & - & - & - & - \\
    ~ & 135 & 1.9 & - & - & - & - & - & - \\
    ~ & 192 & 2.2 & - & - & - & - & - & - \\
    ~ & 192 & 2.1 & - & - & - & - & - & - \\
    ~ & 84 & 3.0 & - & - & - & - & - & - \\
    ~ & 1998 & 2.0 & - & - & - & - & - & - \\
    ~ & 207 & 2.1 & - & - & - & - & - & + \\
    ~ & 81 & 3.3 & - & - & - & - & - & + \\
    ~ & 189 & 1.9 & - & - & - & - & - & + \\
    ~ & 72 & 5.3 & - & - & - & - & - & + \\
    ~ & 189 & 2.3 & - & - & - & - & - & + \\ \hline
  \end{tabular}
  \end{tiny}
  \label{tab:gap-filling-efficiency-tandem-repeats}
\end{table}

\subsection*{Time and memory usage}

\begin{figure}[!h]
  \includegraphics[scale=0.33,angle=0]{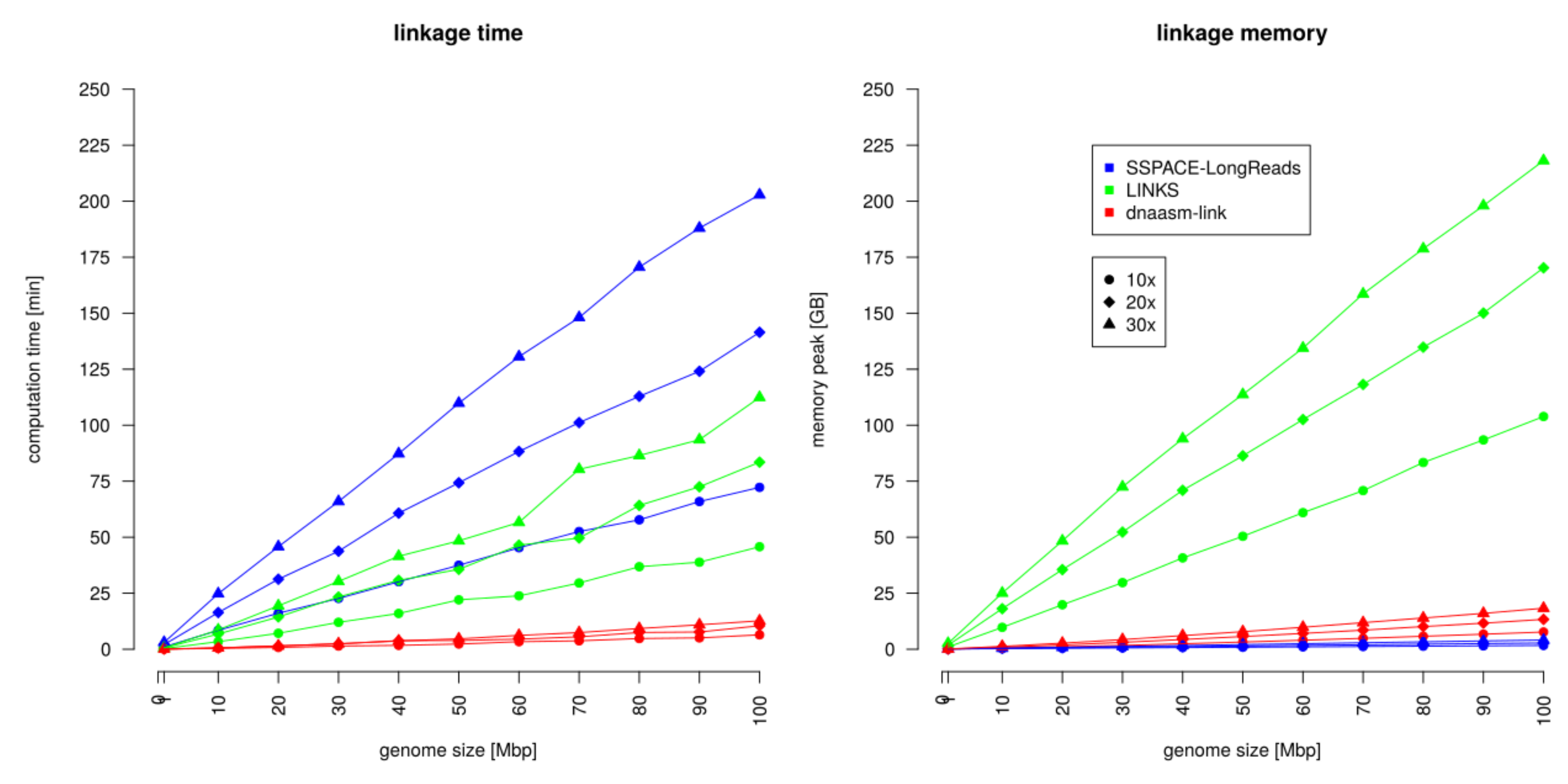}
  \caption{Comparison of calculation time and the peak of RAM memory usage of SSPACE-LongReads, LINKS and dnaasm-link applications. The experiment was carried out on the \textit{Caenorhabditis elegans} genome (GenBank NC\_003279 ... NC\_003284, NC\_001328). Firstly, a set of eleven sub-genomes of sizes 1Mbp, 10Mbp, 20Mbp ... 100Mbp was generated from previously mentioned genome. Then, for each sequence a set of long and short DNA reads was generated, short DNA reads were \textit{de novo} assembled by ABySS tool. Finally, set of resultant contigs and long DNA reads were used as input data sets in the presented experiment.}
  \label{fig:linkage_time_memory}
\end{figure}

We examined dnaasm-link application in terms of performance, which can be crucial in the analysis of large volume sequencing data.
Our application was compared with LINKS~\cite{links} and SSPACE-LongRead~\cite{sspace-longread} in terms of time and memory usage.
The results of the experiment are presented in Fig~\ref{fig:linkage_time_memory}.

As expected, combining contigs in applications with accurate mapping takes much more time than in k-mer based tools, in particular, because of the duration time of mapping of long DNA reads to preassembled contigs.
For example, the calculation time of SSPACE-LongRead application, for which the BLASR~\cite{blasr} software is used in the mapping process, is over 15 times longer than for tools using a k-mer approach, like dnaasm-link tool.
Our tool is significantly faster than LINKS application, because LINKS, which uses a~similar algorithm, is implemented in Perl.
In addition, the LINKS application requires much more RAM memory;
for example, for a~genome of size 100 Mbp and the coverage of long reads equal to 30x, the LINKS application uses over 200 GB of RAM memory, and our application only 18.3 GB.

\section*{Discussion}

The dnaasm-link is the new tool for both: the connection of contigs and filling the gaps between them with long DNA reads.
The presented results indicate that the application works comparatively with existing tools in terms of the quality of the resultant DNA sequences.
However, the application works significantly faster with much less RAM memory usage, which can be crucial for large volume sequencing data.
Moreover, the presented software contains a module for filling the gaps between contigs by a specified sequence from an appropriate long DNA read, which is not implemented in similar tools.

The procedure of filling the gaps through the appropriate fragment of a specified long DNA read can significantly increase the parameters of the resulting DNA sequences.
In the presented study we indicated that a very large number of complex DNA structures, especially tandem repeats, could not be properly reproduced without using long DNA reads.
Moreover, the addition of long DNA reads, even with very low coverage, can significantly reduce the number of resultant DNA sequences and improve their consistency in relation to the results obtained only from short DNA reads.

In the presented application, a~gap within scaffolds could be optionally filled with a fragment of a single long DNA read.
However, this solution is not ideal, because such a read may contain many errors, especially, if the long reads are raw - errors have not been corrected before.
In order to control this issue, in the future we plan to add a module to create consensus from several DNA reads.
The result of the consensus of several long reads would be inserted into the gap instead of the raw fragment of a single long read, which would significantly reduce the number of errors in the considered DNA fragments.
However, the preliminary study shows a big increase in time complexity, when consensus is calculated with the use of multi-alignment dynamic programming algorithm.

In the future, we also plan to add a module for the analysis of the similarity of k-mers, which would take into account the fact that the k-mers may contain errors.
The presented tool is based on k-mers, which should contain as few errors as possible, because each single error in the specified DNA sequence causes the creation of $k$ erroneous k-mers in k-spectrum.
To deal with this problem, in the next version of the software we will add a module which will investigate a profile of specified k-mer and compare it to the profiles of the another k-mers.
The profile mentioned will contain several information, e.g. number of specified 2-mers and their location in the investigated k-mer.

The presented application is available under GNU Library or Lesser General Public License version 3.0 (LGPLv3).
In order to easily use the software, the demo application with web interface as well as Docker~\cite{docker} container with dnaasm-link tool are available.
What is more, the user can download binary files as well as source code and compile the application with any changes in the algorithm.
The correspondent links, additional data and more information can be found at project homepage http://dnaasm.sourceforge.net.

\section*{Conclusions}
As more and more genomes are sequenced, it becomes desirable to correctly reproduce their DNA sequences, especially, from short and long DNA reads.
Here we have presented dnaasm-link, a tool for linking contigs, a result of \textit{de novo} assembly of second-generation sequencing data, with long DNA reads.

\section*{Availability of data and materials}
dnaasm-link is implemented in C++, and is freely available under GNU Library or Lesser General Public License version 3.0 (LGPLv3).
It and related materials can be downloaded from project homepage http://dnaasm.sourceforge.net.

\section*{Funding}
This work was supported by the statutory research of Institute of Computer Science of Warsaw University of Technology.

\section*{Authors’ contributions}
RN identified the problem, RN, WF and WK designed the approach.
WF implemented the software.
WF and WK worked on testing and validation, WK and RN wrote the manuscript.
All authors read and approved the final manuscript.

\bibliographystyle{plain}
\bibliography{dnaasm-link_BMC_Bioinformatics_2018_manuscript}

%\todo{mismatches and indels for E.coli counted from first 40000 reads, which sum equal to 68364479 bp}

\end{document}